\documentclass{article}%
\usepackage{amsmath}
\usepackage{amsfonts}
\usepackage{amssymb}
\usepackage{graphicx}%
\begin{document}

\title{The difference of boundary effects between Bose and Fermi systems}
\author{Hai Pang$^{1}$, Wu-Sheng Dai$^{1,2}$ \thanks{ daiwusheng@tju.edu.cn}, Mi
Xie$^{1,2}$ \thanks{ xiemi@tju.edu.cn}\\{\footnotesize $^{1}$Department of Physics, Tianjin University, Tianjin
300072, P. R. China}\\{\footnotesize $^{2}$LiuHui Center for Applied Mathematics, Nankai University
\& Tianjin University, Tianjin 300072, P. R. China}}
\date{}
\maketitle

\begin{abstract}
In this paper, we show that there exists an essential difference of boundary
effects between Bose and Fermi systems both for Dirichlet and Neumann boundary
conditions: at low temperatures and high densities the influence of the
boundary on the Bose system depends on the temperature but is independent of
the density, but for the Fermi case the influence of the boundary is
independent of the temperature but depends on the density, after omitting the
negligible high-order corrections. We also show that at high temperatures and
low densities the difference of the influence of the boundary between Bose and
Fermi systems appears in the next-to-leading order boundary contribution, and
the leading boundary contribution is independent of the density. Moreover, for
calculating the boundary effects at high temperatures and low densities, since
the existence of the boundary modification causes the standard virial
expansion to be invalid, we introduce a modified virial expansion.

\end{abstract}

\vskip 0.5cm


\section{Introduction}

In small-size and low-temperature systems, the mean thermal wavelength of
particles is comparable with the size of the system, and the boundary effect
becomes one of the most important quantum effects. The influence of boundaries
is widely discussed. Based on Maxwell-Boltzmann statistics, the boundary
effect on the classical ideal gas confined in a narrow box or in spherical and
cylindrical geometries is studied theoretically in \cite{casimir1,casimir2}.
For quantum gases, the influence of the boundary on Bose gases is discussed in
\cite{Ziff}; the shape and topology dependence of the boundary effect on ideal
quantum gases confined in irregular containers is discussed in \cite{gkac};
some exact and approximative solutions for quantum gases in finite-size
containers are provided in \cite{mac}; a general result for the boundary
modification on ideal quantum gases in arbitrary dimensions is provided in
\cite{hsgas}. In Bose systems, the choice of the boundary condition may play
an important role. Especially, for Bose systems it may display a phase
transition --- the Bose-Einstein condensation. Systematic discussions on this
subject are given in \cite{Robinson,VVZ,LW,BZ}. Moreover, some experimental
studies (e.g., \cite{1,2,3,nanotube1,nanotube2}) show that the influence of
the boundary may be remarkable in low-temperature quantum gases confined in
small volumes.

In the present paper we pay special attention to two-dimensional systems. Many
novel properties of the gases absorbed within a bundle of carbon nanotubes are
reported \cite{RMP}, and, especially, under appropriate thermodynamic
conditions gases adsorbed within the nanotubes and on the external surface of
the bundle will display two-dimensional behaviour
\cite{StanCole1998,dresselhaus1999,gatica2001}. Moreover, we also provide some
three-dimensional results.

Different boundary conditions lead to different boundary effects. Two kinds of
boundary conditions are considered: the Dirichlet boundary condition and the
Neumann boundary condition.

The primary motivation of this work is to compare the boundary effects on Bose
and Fermi systems. The analysis given by \cite{mac} shows that the ideal
quantum gas can be used as an efficient tool for studying the boundary effects
in many quantum systems, including electron gases, interacting Bose gases,
etc, since in an interacting system there are two kinds of effects: one arises
from the classical interaction, e.g., the van der Waals interaction, and the
other is the quantum effect. In ideal quantum gases though there are no
classical interactions the quantum effects, such as the exchange interaction,
remain. The boundary effect is, however, a kind of quantum effect, which
becomes important when the mean thermal wavelength of particles is comparable
with the size of the system, so the boundary effects on ideal and interacting
gases are similar and the conclusions of the ideal gases are of widespread applicability.

In finite volumes (areas), energy levels are discrete, and the spectra of the
particles are shape dependent and sensitive to the topology \cite{gkac}. To
take boundary effects into account, strictly speaking, we need to perform the
summation over all possible states directly, but that is in general hardly to
be solved. In \cite{gkac} and \cite{mac}, some methods are developed to
perform the summation. Based on a mathematical work given by M. Kac
\cite{kac}, a method for calculating the boundary modification to the ideal
quantum gases in irregular-shaped containers is provided in \cite{gkac}; some
thermodynamic quantities for two- and three-dimensional systems are calculated
in \cite{mac} and some of the results are exact solutions. Different boundary
conditions lead to different boundary effects. The equation of state for ideal
quantum gases with boundary modifications is provided in \cite{gkac} and
\cite{mac}. The boundary condition used in these two papers is the Dirichlet
boundary condition. In the following, we will consider the boundary effects
corresponding to two kinds of boundary conditions: the Dirichlet boundary
condition and the Neumann boundary condition. It is straightforward to obtain
the equation of state with the Neumann boundary condition by use of the method
given in \cite{gkac} and \cite{mac}. The equation of state of Fermi and Bose
gases in confined space with Dirichlet and Neumann boundary conditions can be
expressed as%
\begin{equation}
\frac{P\Omega}{kT}=\sum\limits_{\sigma}A_{\sigma}h_{\sigma}(z), \label{1}%
\end{equation}%
\begin{equation}
N=\sum\limits_{\sigma}A_{\sigma}h_{\sigma-1}(z), \label{2}%
\end{equation}
where $h_{\sigma}\left(  z\right)  $ equals the Bose-Einstein integral
$g_{\sigma}\left(  z\right)  $ or the Fermi-Dirac integral $f_{\sigma}\left(
z\right)  $ in the Bose or the Fermi case, respectively:
\begin{equation}
h_{\sigma}(z)=\left\{
\begin{array}
[c]{l}%
\displaystyle g_{\sigma}\left(  z\right)  =\frac{1}{\Gamma(\sigma)}\int
_{0}^{\infty}\frac{x^{\sigma-1}dx}{z^{-1}e^{x}-1}=\sum\limits_{l=1}^{\infty
}\frac{z^{l}}{l^{\sigma}},\\
\displaystyle f_{\sigma}\left(  z\right)  =\frac{1}{\Gamma(\sigma)}\int
_{0}^{\infty}\frac{x^{\sigma-1}dx}{z^{-1}e^{x}+1}=\sum\limits_{l=1}^{\infty
}\left(  -1\right)  ^{l-1}\frac{z^{l}}{l^{\sigma}}.
\end{array}
\right.  \label{005}%
\end{equation}
In these equations and the following, the upper sign stands for the Dirichlet
boundary condition and the lower sign for the Neumann boundary condition. In
two dimensions, $\Omega=S$ is the area, $\sigma=2,3/2,1$, and $A_{2}%
=gS/\lambda^{2}$, $A_{3/2}=\mp\left(  1/4\right)  gL/\lambda$, $A_{1}=g\chi
/6$, where $\lambda=h/\sqrt{2\pi mkT}$ is the mean thermal wavelength, $L$ is
the perimeter of the boundary, $\chi$ is the Euler-Poincar\'{e} characteristic
number (the Euler-Poincar\'{e} characteristic number reflects the connectivity
--- a topological property --- of a two-dimensional system), and $g$ denotes
the number of internal degrees of freedom, the degeneracy number of spin
states (for bosons we take $g=1$). For the case of the gas confined in a
three-dimensional box, $\Omega=V$ is the volume, $\sigma=5/2,2,3/2,1$, and
$A_{5/2}=gV/\lambda^{3}$, $A_{2}=\mp\left(  1/4\right)  gS/\lambda^{2}$,
$A_{3/2}=\left(  1/16\right)  gL/\lambda$, $A_{1}=\mp\left(  1/8\right)  g$,
where $S$ is the area of the surface of the box, $L=4(L_{x}+L_{y}+L_{z})$ is
the total length of the sides of the box.

The difference between the Dirichlet case and Neumann case is that the leading
contributions of these two kinds of boundaries have opposite signs: the
contribution of the Dirichlet boundary is negative, and the contribution of
the Neumann boundary is positive.

The standard virial expansion approach cannot be used due to the existence of
the boundary modification. For analysing the boundary effect in
high-temperature and low-density systems, we introduce a modified virial expansion.

In this paper, we compare the boundary effects on the Bose system and on the
Fermi system. At low temperatures and high densities, our results will show
that the boundary effects on Bose and Fermi systems are different essentially:
after omitting the negligible high-order corrections, for Bose systems the
influence of the boundary depends on the temperature but is independent of the
density; however, contrary to the Bose case, the influence of the boundary on
Fermi systems depends on the density but is independent of the temperature. At
high temperatures and low densities, the difference of the boundary effects
between Bose and Fermi systems appears in the next-to-leading boundary
contributions: the next-to-leading boundary contributions to Bose systems and
to Fermi systems have opposite signs; they are functions of both temperature
and density, though the leading boundary contribution is independent of the density.

In Section 2, we compare and discuss the difference of boundary effects
between Bose and Fermi systems both for Dirichlet and Neumann boundary
conditions. In Section 3, by introducing a modified virial expansion, we
calculate the boundary effects on Bose and Fermi systems at high temperatures
and low densities. The conclusions are summarized in Section 4.

\section{The difference between Bose and Fermi systems at low temperatures and
high densities}

In this section, we compare the boundary effects on the two-dimensional Bose
and Fermi systems at low temperatures and high densities. We show that the
boundary effects on Bose and Fermi systems are different essentially, and
discuss the reason of such a difference.

We will take the specific heat, a quantity directly accessible to experimental
measurement, as an example, and our main conclusion holds also for other
thermodynamic quantities.

\textit{The Bose case. }For Bose cases, from the equation of state, Eqs.
(\ref{1}) and (\ref{2}), we can obtain the specific heat at low temperatures
and high densities. From Eq. (\ref{2}) one can determine $\alpha=\left[
1\mp\left(  \sqrt{\pi}/4\right)  \left(  L\lambda/S\right)  e^{n\lambda^{2}%
/2}\right]  e^{-n\lambda^{2}}$ approximately, where $\alpha=-\mu/\left(
kT\right)  $, with $\mu$ the chemical potential, and $n=N/S$ is the mean
number density, where $N$ is the expectation value of the particle number in
the grand canonical ensemble. Substituting $\alpha$ into Eq. (\ref{1}), we
obtain the grand potential. Starting from the grand potential we can calculate
the specific heat,
\begin{equation}
\frac{C_{V}}{k}=\frac{S}{\lambda^{2}}\left\{  \frac{\pi^{2}}{3}-\left[
\left(  n\lambda^{2}\right)  ^{2}+2n\lambda^{2}+2\right]  e^{-n\lambda^{2}%
}\right\}  \mp\frac{L}{8\lambda}\left\{  \frac{3}{2}\zeta\left(  \frac{3}%
{2}\right)  -\frac{\sqrt{\pi}}{2}\left[  \left(  n\lambda^{2}\right)
^{2}+3n\lambda^{2}+6\right]  e^{-n\lambda^{2}/2}\right\}  . \label{16}%
\end{equation}
In this section, we always drop out of the topological terms which are often
negligible. The terms proportional to $S/\lambda^{2}$ in Eq. (\ref{16}) are
the specific heat in free space, and the remaining terms are the boundary modifications.

The percentage of the boundary modification to specific heat is%
\begin{align}
\frac{C_{V}-C_{V}^{0}}{C_{V}^{0}}  &  =\mp0.149\frac{L}{\sqrt{S}}\frac
{\lambda}{\sqrt{S}}\left[  1-0.226\left(  n\lambda^{2}\right)  ^{2}%
e^{-n\lambda^{2}/2}+\cdots\right] \nonumber\\
&  \approx\mp0.149\frac{L}{\sqrt{S}}\frac{\lambda}{\sqrt{S}}\propto\frac
{1}{\sqrt{T}}.
\end{align}
We can see that the boundary effect on a Bose system, after omitting the
negligible high-order corrections, is independent of the density; it is
proportional to $\lambda/\sqrt{S}$, which is the ratio between the thermal
wavelength and the linear size of the container, and $L/\sqrt{S}$, which
reflects to some extent the information of the shape of the boundary. We can
see that as the temperature falls, the contribution from the boundary will
become more and more important.

The above result is obtained in low temperatures. It should be emphasized that
this result will become invalid when the fugacity $z\rightarrow1$, since the
result which we based on given by Refs. \cite{gkac} and \cite{mac} is invalid
when $z\rightarrow1$. If the system can display Bose-Einstein condensation,
the case $z\rightarrow1$ corresponds to a non-zero temperature $T_{c\text{.}}$
That is to say, if the Bose-Einstein condensation occurs, our result will be
invalid at a certain non-zero temperature. However, in the above case (the
two-dimensional ideal Bose gas with Dirichlet and Neumann boundaries) there is
no Bose-Einstein condensation \cite{Robinson}. This means that only when
$T\rightarrow0$ the fugacity $z\rightarrow1$. Nevertheless, the modification
to the equation of state, Eqs. (\ref{1}) and (\ref{2}), will be valid only
when the boundary contribution is small \cite{kac}. Consequently, for the Bose
case there exists a lower limit on temperatures in the range of the
applicability of the above result. This lowest temperature can be roughly
estimated: the boundary modification should be less than the leading
contribution. From Eq. (\ref{16}) we can obtain this lower limit for a square
two-dimensional system directly: the above result is valid only when the
temperature $T>0.057h^{2}/\left(  mka^{2}\right)  $, where $a$ is the side
length of the system. For illustration, such a temperature for $^{23}$Na\ with
size $50nm$ is about $2\times10^{-5}K$. Moreover, the boundary influence on
the condensation is a very important and interesting subject, which has been
considered in \cite{Robinson,VVZ,LW,BZ}.

\textit{The Fermi case. }For Fermi cases, the specific heat is%
\begin{equation}
\frac{C_{V}}{Nk}=\frac{\pi^{2}}{3}\frac{kT}{\varepsilon_{F}^{0}}\left[
\left(  1\mp\frac{1}{2}\delta\right)  \mp\frac{7\pi^{2}}{160}\delta\left(
\frac{kT}{\varepsilon_{F}^{0}}\right)  ^{2}+\cdots\right]  .
\end{equation}
The ratio of the boundary modification to the specific heat in free space is%
\begin{equation}
\frac{C_{V}-C_{V}^{0}}{C_{V}^{0}}=\mp\frac{1}{2}\delta\left[  1+\frac{7\pi
^{2}}{80}\left(  \frac{kT}{\varepsilon_{F}^{0}}\right)  ^{2}+\cdots\right]
\sim\mp\frac{1}{2}\delta, \label{Cvfermi2}%
\end{equation}
where
\[
\delta=\frac{1}{2}\sqrt{\frac{g}{\pi}}\frac{L}{\sqrt{S}}\frac{1}{\sqrt{S}%
}\sqrt{\frac{S}{N}}.
\]
Unlike the Bose case, after omitting the contribution suppressed by $\left(
kT/\varepsilon_{F}^{0}\right)  ^{2}$, the ratio $(C_{V}-C_{V}^{0})/C_{V}%
^{0}\propto\mp L/\sqrt{S}(1/\sqrt{S})\sqrt{S/N}$, is independent of the
temperature but is determined by such three factors: the number density $N/S$
(or, the mean space between particles $\sqrt{S/N}$), the shape, which is
described by $L/\sqrt{S}$ to some extent, and the linear size $\sqrt{S}$. In
other words, the boundary effects on two-dimensional Fermi systems depend on
the density, the size, and the geometrical properties of the system, but is
independent of the temperature.

\textit{The difference between Bose and Fermi systems. }In a word, in addition
to the size and shape, the boundary effect on Bose systems depends almost only
on the temperature, but that on Fermi systems depends almost only on the
density. Such a difference is quite essential. In the following we will
analyse the reason.

Essentially, the influence of the boundary is determined by the thermal
wavelength of the particles and the size of the system. The lower the energy
of a particle, the longer is the wavelength and then the stronger is the
boundary effect. Therefore, the particles in lower energy levels will be
influenced more strongly than those in higher energy levels.

First, we discuss the influence of the temperature on the boundary effects.

For a Bose system, when the temperature falls, the particles will tend to
occupy lower energy levels, so the boundary effect will become stronger. As a
result, the boundary effect will be a function of the temperature.

For a Fermi system, there exists the Fermi energy. At low temperatures almost
all the particles are in the states below the Fermi energy. The magnitude of
the influence of the boundary on the particles is determined by the
wavelength: the shorter the wavelength the weaker the boundary influence is.
Generally speaking, the energy of the particles near the Fermi surface is
relatively high, so the wavelength is short. Therefore in a low-temperature
system the contribution of the boundary influence mainly comes from the
particles far below the Fermi surface. At very low temperatures, a small
change in the temperature only influences the particles near the Fermi
surface; however, due to their short wavelengths, the boundary contribution
from such particles is very small compared with the whole boundary effect.
Therefore, in low-temperature Fermi systems the boundary effect is almost
independent of the temperature.

Next, we discuss the influence of the density on the boundary effects.

For a Bose system with fixed size and shape, the change in the density will
change the total number of particles, but almost has no influence on the
relative distribution of particles. What we consider is the ratio between the
boundary contribution and the result of free space, i.e., the percentage of
the boundary contribution, which mainly depends on the distribution of the
particles. Therefore, such a ratio will not change after changes in density.

For a Fermi system with fixed size and shape, the ratio of the boundary
contribution to the free space result will get smaller as the density
increases. When the volume is given, as the density increases, the number of
particles increases. Roughly speaking, the energy of the newly added particle
will be higher than the original Fermi energy, i.e., the wavelength of the
newly added particle will be shorter than the original mean wavelength. As
mentioned above, the boundary effect is mainly determined by the magnitude of
the wavelength of the low-energy particles. Therefore with the density
increasing the proportion of the influence of the boundary with respect to the
result in free space will be suppressed.

Particularly, it should be emphasized that in the Fermi case the influences on
different thermodynamic quantities are different: for example, for the
Dirichlet boundary condition, the influence on the specific heat is negative,
but on the Fermi energy is positive. For illustrating this, we give the
chemical potential and Fermi energy in a two-dimensional system,
\begin{equation}
\mu=\varepsilon_{F}^{0}\left[  \left(  1\pm\delta\right)  \mp\delta\frac
{\pi^{2}}{24}\left(  \frac{kT}{\varepsilon_{F}^{0}}\right)  ^{2}%
+\cdots\right]  , \label{cp2}%
\end{equation}

\begin{equation}
\varepsilon_{F}=\varepsilon_{F}^{0}\left(  1\pm\delta\right)  ,
\end{equation}
where $\varepsilon_{F}^{0}=\left(  \hbar^{2}/2m\right)  \left(  4\pi
N/gS\right)  $ is the Fermi energy in two-dimensional free space, and $\left(
\varepsilon_{F}-\varepsilon_{F}^{0}\right)  /\varepsilon_{F}^{0}=\pm\delta$.
Note that this boundary modification to the case of the Dirichlet boundary
condition is positive and to the case of the Neumann boundary condition is
negative; however, for the specific heat, Eq. (\ref{Cvfermi2}), the
modification to the Dirichlet case is negative, but to the Neumann case is positive.

Moreover, the boundary effect on the chemical potential deserves special
attention. For illustrating this, we give the three-dimensional chemical
potential and Fermi energy,%
\begin{equation}
\mu_{3}=\varepsilon_{F3}^{0}\left[  \left(  1\pm\Delta\right)  -\frac{\pi^{2}%
}{12}\left(  \frac{kT}{\varepsilon_{F3}^{0}}\right)  ^{2}+\cdots\right]  ,
\label{cp3}%
\end{equation}
and%
\begin{equation}
\varepsilon_{F3}=\varepsilon_{F3}^{0}\left(  1\pm\Delta\right)  ,
\end{equation}
where $\varepsilon_{F3}^{0}=\left(  \hbar^{2}/2m\right)  \left(  6\pi
^{2}N/gV\right)  ^{2/3}$ is the Fermi energy in three-dimensional free space
and
\[
\Delta=\frac{1}{4}\left(  \frac{\pi g}{6}\right)  ^{1/3}\frac{S}{V^{2/3}}%
\frac{1}{V^{1/3}}\left(  \frac{V}{N}\right)  ^{1/3}.
\]
The second-order term of the two-dimensional chemical potential (\ref{cp2})
only includes the boundary contribution, while the second-order term of the
three-dimensional chemical potential (\ref{cp3}) is independent of the
boundary, i.e., there is no second-order boundary correction to the
three-dimensional chemical potential. This leads to such a result: in two
dimensions, the second-order contribution to the ratio $\left(  \mu-\mu
^{0}\right)  /\mu^{0}=\pm\delta\left[  1-\left(  \pi^{2}/24\right)  \left(
kT/\varepsilon_{F}^{0}\right)  ^{2}+\cdots\right]  $ is negative, but in three
dimensions the second-order contribution to the ratio $\left(  \mu_{3}-\mu
_{3}^{0}\right)  /\mu_{3}^{0}=\pm\Delta\left[  1+\left(  \pi^{2}/12\right)
\left(  kT/\varepsilon_{F3}^{0}\right)  ^{2}+\cdots\right]  $ is positive. The
reason is that in the result of the standard statistical mechanics, which is
under the thermodynamic limit approximation, the two-dimensional chemical
potential has no second-order term \cite{Ann.}, so the second-order
contribution in Eq. (\ref{cp2}) only comes from the boundary effect. In
three-dimensional free space, however, the chemical potential does have the
second-order term, but the three-dimensional boundary correction, which
behaves like the two-dimensional chemical potential, has no second-order term.

\section{Boundary effects at high temperatures and low densities: a modified
virial expansion}

The existence of the boundary modification causes the standard virial
expansion to be invalid. For analysing the boundary effect at high
temperatures and low densities, we introduce a modified virial expansion approach.

In statistical mechanics, in the limit $T\rightarrow\infty$, hence
$\lambda\rightarrow0$, all the thermodynamic quantities can be expanded with
respect to $n\lambda^{2}$ since $n\lambda^{2}=N\lambda^{2}/S\ll1$, which is
the so-called virial expansion. The main idea of the virial expansion in two
dimensions is to suppose that the equation of state can be expanded as
$PS/(NkT)=\mathop{\displaystyle\sum}\limits_{l=1}^{\infty}a_{l}\left(
n\lambda^{2}\right)  ^{l-1}$, where the coefficients $a_{l}$\ are referred to
as the virial coefficients. Nevertheless, since Eqs. (\ref{1}) and (\ref{2})
contain the terms which describe the influence of the boundary shape and
topology, the expansion parameter of the virial expansion $n\lambda^{2}$ (or
$N\lambda^{2}/S$) is not the unique expansion parameter, so the thermodynamic
quantities cannot be expressed as the series of $n\lambda^{2}$. In other
words, it is impossible to perform the standard virial expansion.

For calculating the thermodynamic quantities at high temperatures and low
densities, we introduce a new approach which is, in fact, a modified virial
expansion. We suppose that the equation of state can be expressed as a series,%
\begin{equation}
\frac{PS}{NkT}=\sum\limits_{l=1}^{\infty}a_{\left(  l+1\right)  /2}\left(
n\lambda^{2}\right)  ^{\left(  l-1\right)  /2}, \label{4}%
\end{equation}
where $a_{\left(  l+1\right)  /2}$\ are the modified virial coefficients.

Using Eqs. (\ref{2}) and (\ref{005}), we obtain%
\begin{equation}
n\lambda^{2}=\left(  z+\eta\frac{z^{2}}{2}+\cdots\right)  \mp\frac{1}{4}%
\frac{L\lambda}{S}\left(  z+\eta\frac{z^{2}}{\sqrt{2}}+\cdots\right)
+\frac{\chi}{6}\frac{\lambda^{2}}{S}\left(  z+\eta z^{2}+\cdots\right)  .
\label{5}%
\end{equation}
In this equation and following, for the Bose case $\eta=+1$ and for the Fermi
case $\eta=-1$. In this equation, the expansion parameters include
$N\lambda^{2}/S$, $L\lambda/S$,\textbf{\ }$\lambda^{3}/\left(  LS\right)  $
and $(\lambda^{2}/S)\chi/6$. Therefore we write $z$ in the form
\begin{equation}
z=n\lambda^{2}\sum\nolimits_{l=1}^{\infty}b_{\left(  l+1\right)  /2}\left(
n\lambda^{2}\right)  ^{\left(  l-1\right)  /2},
\end{equation}
where $b_{\left(  l+1\right)  /2}$ are the coefficients of the expansion. Then
we can obtain an equation of $\lambda$. By equating the coefficients of each
power of $\lambda$, we can obtain the coefficients $b_{\left(  l+1\right)
/2}$ and then obtain the modified virial coefficients $a_{\left(  l+1\right)
/2}$:%
\begin{equation}
a_{1}=1,a_{3/2}=0,a_{2}=-0.25\eta,a_{5/2}=\mp0.037\eta\frac{1}{\sqrt{N}}%
\frac{L}{\sqrt{S}},a_{3}=0.028-0.003\eta\frac{1}{N}\frac{L^{2}}{S},\cdots.
\end{equation}

One can see that $a_{\left(  l+1\right)  /2}$ are related to the geometry of
the system, and if the thermodynamic limit approximation is taken, they will
return to the common virial coefficients. By $a_{\left(  l+1\right)  /2}$, we
can express the expansions for the thermodynamic quantities. As an example, we
give the expansions of chemical potential and specific heat:%
\begin{align}
\frac{\mu}{kT}  &  =\ln\left(  n\lambda^{2}\right)  \pm0.25\frac{L\lambda}%
{S}+0.031\left(  \frac{L\lambda}{S}\right)  ^{2}-0.167\frac{\chi\lambda^{2}%
}{S}+\cdots\nonumber\\
&  -0.5\eta n\lambda^{2}\left[  1\pm0.146\frac{L\lambda}{S}+0.011\left(
\frac{L\lambda}{S}\right)  ^{2}+\cdots\right]  +0.042\left(  n\lambda
^{2}\right)  ^{2}\left[  1\pm0.100\frac{L\lambda}{S}+\cdots\right]
+\cdots,\nonumber
\end{align}

\begin{align}
\frac{C_{V}}{Nk}  &  =\left[  1\pm0.063\frac{L\lambda}{S}+\cdots\right]
+0.25\eta n\lambda^{2}\left[  \pm0.110\frac{L\lambda}{S}+0.021\left(
\frac{L\lambda}{S}\right)  ^{2}+\cdots\right] \nonumber\\
&  -0.028\left(  n\lambda^{2}\right)  ^{2}\left[  1\pm0.188\frac{L\lambda}%
{S}+\cdots\right]  .\nonumber
\end{align}

In comparison with the virial expansion, one can see that the boundary effects
are reflected mainly by the terms which are proportional to $L\lambda/S$ and
its powers.

The result shows that at high temperatures and low densities, both for Bose
and Fermi cases, the leading contribution of the boundary to the chemical
potential and the specific heat is a function of temperature and geometrical
property of the system, but is independent of the density of the system. In
the next-to-leading order boundary contribution, the difference between Bose
and Fermi systems appears: the modifications to Bose systems and to Fermi
systems have opposite signs. Moreover, the boundary modifications to different
thermodynamic quantities are different. One can also verify that the
properties we have drawn in this section are also satisfied by the other
thermodynamic quantities.

Note that the same procedures can be used to the three-dimensional case
directly by expressing the equation of state and the fugacity $z$ as:%
\begin{equation}
\frac{PV}{NkT}=\sum\limits_{l=1}^{\infty}c_{\left(  l+2\right)  /3}\left(
n\lambda^{3}\right)  ^{\left(  l-1\right)  /3}\text{\ \ and \ \ }%
z=n\lambda^{3}\sum\limits_{l=1}^{\infty}d_{\left(  l+2\right)  /3}\left(
n\lambda^{3}\right)  ^{\left(  l-1\right)  /3}.
\end{equation}

\section{Conclusion}

In conclusion, we show the essential difference of boundary effects on Bose
and Fermi systems in two dimensions: at low temperatures and high densities,
for Bose systems, boundary effects are almost independent of the density but
depend on the temperature; while for Fermi systems, the effects are almost
independent of the temperature but depend on the density. At high temperatures
and low densities, the difference of boundary effects between Bose and Fermi
systems appears in the next-to-leading order boundary contributions: the
next-to-leading order contributions to Bose systems and to Fermi systems have
opposite signs; moreover, the leading boundary contribution is independent of
the density.

Strictly speaking, for a finite-size system one should use the canonical
formalism, but it is shown in \cite{mac} that the result based on the grand
canonical formalism for finite-size systems is valid for most realistic cases.
A detailed discussion for the canonical ensemble formalism is given by
\cite{PZ,ZB}.\ In this paper attention is concentrated on the difference of
boundary effects between Bose and Fermi systems, and we have not considered
the Bose-Einstein condensation since there is no corresponding phase
transition in the Fermi system at low temperatures. Of course, the boundary
effect on the Bose-Einstein condensation is very important. Systematic
discussions on this subject can be found in \cite{Robinson,VVZ,LW,BZ}. Note
that in this paper though our attention focuses mainly on the two-dimensional
cases, it is easy to check that most of our conclusion is also valid for three dimensions.

\vskip0.5cm This work is supported in part by the Science fund of Tianjin
University and Education Council of Tianjin, People's Republic of China, under
Project No. 20040506. The computation of this project was performed on the
HP-SC45 Sigma-X parallel computer of ITP and ICTS, CAS.

\end{document}